\documentstyle[aps,preprint,graphicx]{revtex}

\tighten

\begin{document}
\title{Dynamic charge density correlation function in weakly charged
polyampholyte globules} \author{Hindrik Jan Angerman$^1$ and Eugene
Shakhnovich$^2$\\ {\it $^1$University of Groningen, Department of Polymer Chemistry, Nijenborgh 4, 9747 AG
Groningen, The Netherlands}\\ {\it $^2$Harvard University, Department of Chemistry,
12 Oxford Street, Cambridge MA 02138, USA}} \date{\today}
\maketitle {\abstract We study solutions of statistically neutral
polyampholyte chains containing a large fraction of neutral
monomers. It is known that, even if the quality of the solvent with
respect to the neutral monomers is good, a long chain will collapse
into a globule. For weakly charged chains, the interior of this globule
is semi-dilute. This paper considers mainly $\theta-$solvents, and we
calculate the dynamic charge density correlation function $g(k,t)$ in
the interior of the globules, using the quadratic
approximation to the Martin-Siggia-Rose generating functional. It is
convenient to express the results in terms of dimensionless space and
time variables. Let $\xi$ be the blob size, and let $\tau$ be the
characteristic time scale at the blob level. Define the dimensionless
wave vector $q=\xi k$, and the dimensionless time $s=t/\tau$. We find
that for $q<1$, corresponding to length scales larger than the blob
size, the charge density fluctuations relax according to $g(q,s)\sim
q^2(1-s^{1/2})$ at short times $s<1$, and according to $g(q,s)\sim q^2
s^{-1/2}$ at intermediate times $1<s<q^{-4}$.  We expect these results
to be valid for wave vectors $q>0.1$, where entanglements are
unimportant.  }\\

\section{Introduction}

A polyampholyte is a polymer chain that contains electrically charged
monomers of both signs. Apart from the charged monomers, the chain may
also contain neutral monomers. In this paper we study polyampholytes
in which the neutral monomers are in the majority, so the chains are
only weakly charged. The positive and the negative monomers are
distributed with equal probability and without correlations along the
chain, separated from each other by flexible neutral spacers. The
chains are present in a solvent. We consider two situations; either
the quality of the solvent with respect to the neutral monomers is
good, or the solvent is under $\theta-$conditions. It was shown in \cite{Joanny1} that in
both situations an isolated neutral chain collapses into a non-compact
globule, provided that the chain length exceeds some critical
value. For weakly charged chains the polymer concentration inside the
globule is low. In case of a good solvent, further collapse is
prevented by the second order virial coefficient (also called the
excluded volume parameter), whereas in case of a $\theta-$solvent, further
collapse is prevented by the third order virial coefficient \cite{Joanny1}. The
non-compact semi-dilute globule can be regarded as a dense melt of
blobs. At length scales small compared to the blob size $\xi$, the
electrostatic interaction is weak (relative to the entropy), and the
system is indistinguishable from a neutral non-collapsed polymer
coil. In case of a good solvent, the chain conformation is a
self-avoiding walk, whereas in case of a $\theta-$solvent, it is a random
walk. On these short length scales, the charged monomers occupy random
positions in space, and so the electrostatic interaction is
unscreened. The importance of the electrostatic interaction relative
to the entropy increases if one considers larger and larger length
scales, and the break-even point is at the blob size. At length scales
larger than the blob size the interaction is strong, and the charged
monomers rearrange themselves spatially in order to minimize the
energy. This leads to a screening of the interaction. A side effect of
these rearrangements of the charges is an effective attraction,
causing the chain to appear collapsed on length scales larger than $\xi$. This
physical picture, originating from Ref. \cite{Joanny1}, shows why the
electrostatic screening length $\kappa^{-1}$ has to be equal to the blob size $\xi$,
which is also equal to the screening length of the excluded volume
interactions in case of a good solvent. Since for weakly charged
polyampholytes the blob size is much larger than the average distance
between two charged monomers, Debye-H$\ddot{\rm u}$ckel theory \cite{Debye} is applicable
(see also Ref. \cite{Joanny1}).  If the chain has a net charge, there is the
possibility that the formation of the globular state is prevented by
the electrostatic repulsion, depending on the chain length and the
magnitude of the excess charge. For the "statistically neutral" chains
considered here, the typical excess charge is proportional to the
square root of the chain length. It was shown in Ref. \cite{Gutin} that in this
case the electrostatic repulsion is insufficient to prevent the
formation of the globular state for long enough chains. The only
effect of the presence of a relatively large excess charge is that the
globule becomes elongated.  In Ref. \cite{Joanny2} multi-chain effects were
studied. It was shown that solutions with non-zero concentration start
to phase separate already at very low concentrations, even if the
quality of the solvent with respect to the neutral monomers is
good. The supernatant consists of isolated spherical globules floating
in the solvent. These globules contain either a single neutral chain,
or two chains whose excess charges cancel each other. The semi-dilute
precipitate, which still has a low polymer concentration, is locally
indistinguishable from the interior of the globules. In this paper we
will concentrate on the dense (semi-dilute) phase.

\section{METHOD AND REGION OF APPLICABILITY}

	We study the dynamics of the charge density fluctuations in
the dense phase (precipitate or globule), using the quadratic
approximation to the Lagrange version [5,6] of the
Martin-Siggia-Rose formalism \cite{MSR}. We will follow closely the
techniques of a paper by Fredrickson and Helfand \cite{Fred}. Readers who wish
to follows the details of our calculation are advised to study
Ref. \cite{Fred} first, especially its appendix. First we will determine in
what length scale regime this formalism is expected to give correct results.  There
are three important length scales in the system: the blob size $\xi$, the
hydrodynamic screening length $\xi_{\rm{H}}$, and the entanglement length $\xi_{\rm{e}}$. At
length scales smaller than $\xi$, the thermal energy is larger than the
electrostatic energy, and the system behaves as if it were a neutral
single chain. Since in the past the dynamics of neutral chains has
been studied extensively \cite{Doi}, and we are mainly interested in the
influence of the electrostatic interaction, we will only consider
length scales that are larger than $\xi$. When observed at these scales,
the chain conformation is always a random walk, even if the quality of
the solvent is good. This is due to a screening of the excluded volume
interactions \cite{Doi}. It means that the excluded volume parameter $v$ enters
the physics at large length scales $>\xi$ only by fixing the blob size \cite{Joanny1}.

Another important length scale is the entanglement length $\xi_{\rm{e}}$. 
It is very difficult
to take the influence of entanglements on the dynamics into
account, and we will not make an attempt to do so. This means
that our results will not be valid on length scales large
compared to the entanglement length $\xi_{\rm{e}}$. One can obtain an
estimate for $\xi_{\rm{e}}$ in the following way. In a polymer melt, the
number of monomers $N_{\rm{e}}$ in between two entanglements is in the
range $N_{\rm{e}}\approx50-500$. Since the interior of the polyampholyte globules can
be regarded as a dense melt of blobs, it follows that the
typical entanglement length is given by

\begin{eqnarray}\label{6}
\xi_{\rm e}\sim\xi N_{\rm e}^{1/2}\sim10\xi.
\end{eqnarray}

At length scales that are large compared to $\xi_{\rm{e}}$ it is to be
expected that the dynamics occurs via some kind of reptation, and on
these scales the relaxation of charge density fluctuations probably
becomes very slow (compare with Ref. \cite{Bouchaud}, where systems with short range
interactions are studied). 

Next we discuss the hydrodynamic screening length $\xi_{\rm H}$. At length scales smaller than
$\xi_{\rm{H}}$, the
chain exhibits Zimm-like behavior, whereas at length scales larger
than $\xi_{\rm{H}}$, it exhibits Rouse-like behavior. Since it was shown in Ref. \cite{Fred}
that the quadratic approximation leads to the prediction of Rouse-like
behavior, our analysis is only applicable to length scales larger than
$\xi_{\rm{H}}$. We need an estimate of its value. According to Ref. \cite{Doi},

\begin{eqnarray}\label{1}
\xi_{\rm H}\sim\frac{1}{\rho b^2},
\end{eqnarray}

where $\rho$ is the monomer number density, and $b$ is the statistical segment
length of a monomer (NB in this paper the sign $\sim$ means: has the same
order of magnitude). First consider the case of a $\theta-$solvent. According
to Ref. \cite{Joanny1} we have

\begin{eqnarray}\label{2}
\rho_\theta\sim\frac{f\ell}{b^4}\hspace{1cm}\xi_\theta\sim\frac{b^2}{f\ell},
\end{eqnarray}

where $\ell:=e^2/4\pi \epsilon k_{\rm{B}}T$ is the Bjerrum length ($\ell =0.7 nm$ in water at room temperature), $\epsilon$ is the
electric permittivity of water, $k_{\rm{B}}$ is Boltzmann's constant, $T$ is the
temperature, and $f$ is the fraction of monomers that carries a
charge. Combining Eqs. (1) and (2), we obtain

\begin{eqnarray}\label{3}
\xi_{{\rm H},\theta}\sim\xi_\theta.
\end{eqnarray}

In a good solvent, characterized by excluded volume parameter $v$, we have \cite{Joanny1}

\begin{eqnarray}\label{4}
\rho_{\rm g}\sim\frac{f^2\ell^2}{vb^5}\hspace{1cm}\xi_{\rm g}\sim\left(\frac{vb^5}{f^3\ell^3}\right)^{1/2}.
\end{eqnarray}

Combining Eqs (1), (2) and (4), we obtain

\begin{eqnarray}\label{5}
\frac{\xi_{\rm H,g}}{\xi_{\rm g}}\sim \left(\frac{\rho_\theta}{\rho_{\rm g}}\right)^{1/2}>>1.
\end{eqnarray}

We conclude that under good solvent conditions the hydrodynamic
screening length is larger than the blob size, whereas under
$\theta-$conditions they are approximately equal. 

Finally, we discuss the region of validity of the quadratic approximation. It  entails
expanding the effective Hamiltonian in powers of the charge density
$\psi(\bf{r})$, and retaining only the second order term. This
approximation can only be justified if the typical amplitude of
$\psi(\bf{r})$ is small. At length scales smaller than $\xi$, the
system is strongly fluctuating, and the amplitude of $\psi(\bf{r})$ is
large. At length scales larger than $\xi$, the charge density
fluctuations are strongly suppressed by the electrostatic
interactions, and the amplitude of $\psi(\bf{r})$ is small. It follows
that the quadratic approximation is acceptable at length scales larger
than $\xi$, but not at length scales shorter than $\xi$. 

Summarizing,
the approximations made in this paper are expected to be reasonable on
length scales that are large compared to both $\xi$ and
$\xi_{\rm{H}}$, and small compared to $\xi_{\rm{e}}$. In order to
ensure that our region of applicability is not empty, we will
henceforth assume that the solvent is under $\theta-$conditions, in
which case $\xi_{\rm{H}}\sim \xi$, and $\xi_{\rm{e}}\sim 10\xi$.

Since in a $\theta-$solution the hydrodynamic interaction is screened
beyond length scale $\xi$, its influence enters the physics on
intermediate length scales $\xi<L<\xi_{\rm{e}}$ in a trivial way via
just one parameter, which is the characteristic time scale $\tau$ at
the blob level. Since inside the blobs the influence of the
electrostatic interactions is negligible and the chain is not
collapsed, $\tau$ can be found by considering a single neutral ideal
chain with hydrodynamic interaction (Zimm-model). This leads to the
estimate \cite{Doi}

\begin{eqnarray}\label{7}
\tau\sim\frac{\eta\xi^3}{k_{\rm B}T}\hspace{1cm}\xi\sim\frac{b^2}{f\ell},
\end{eqnarray}

where $\eta$  is the viscosity of water.

\section{CALCULATION OF THE DYNAMIC CHARGE CORRELATION FUNCTION}

As mentioned in the introduction, we wish to describe the
(semi-dilute) precipitate of a solution of weakly charged
polyampholytes. The individual chains are assumed to be much longer
than the minimum length required for an isolated chain to collapse
into a globule. All chains have exactly the same number of charged
monomers, and these monomers are placed at regular distances along the
backbone. The last two assumptions, which are made in order to
simplify the model, do not restrict the generality of our results. Let
$N$ be the number of charged monomers per chain, $e$ the charge per
monomer, $n_{\rm{p}}$ the number of chains in the precipitate, and $V$
the volume of the precipitate. Note that $V$ cannot be chosen
independently from $n_{\rm{p}}$ and $N$, because the precipitate has a
well-defined density \cite{Joanny1}. Let the coarse grained
conformation of chain $a$ be described by the 3-dimensional vector
$\bf{R}_{\it{a}}(\tau)$ , where $\tau$ is a continuous parameter
running along the backbone. It is defined such that for two
charged monomers neighboring along the chain we have $\Delta \tau=1$. We simplify the
model by smearing out the electric charges along the chain, in such a
way that $e\theta (\tau)d\tau$ is the amount of charge in between the
points labeled by $\tau$ and $\tau+d\tau$, where $\theta(\tau)$ is a
Gaussian random variable with first two moments

\begin{eqnarray}\label{8}
\left<\theta_a(\tau)\right>=0,\hspace{1cm}\left<\theta_a(\tau)\theta_b(\tau')\right>=\delta(\tau-\tau')\delta_{ab}.
\end{eqnarray}

Since we will work in the quadratic approximation, the fact that we
switch to a Gaussian charge distribution has no effect on the final
result. We will describe the system by means of the following
effective Hamiltonian:

\begin{eqnarray}\label{9}
\frac{{\cal H}}{k_{\rm B}T}=\frac{3}{2\tilde{b}^2}\int_0^N{\rm d}\tau\sum_{a=1}^{n_{\rm p}}
\left(\frac{{\rm d}{\bf R}_a(\tau)}{{\rm d}\tau}\right)^2 +\frac{\ell}{2}\sum_{a,b=1}^{n_{\rm p}}
\int_0^N{\rm d}\tau\int_0^N{\rm d}\tau'\frac{\theta_a(\tau)\theta_b(\tau')}{|{\bf R}_a(\tau)-{\bf R}_b(\tau')|},
\end{eqnarray}

where $\tilde{b}$ is the root-mean-square distance between two charged monomers
neighboring along the chain. From now on, we will choose the units of
length, mass and time such that

\begin{eqnarray}\label{10}
\gamma=1,\hspace{1.5cm}k_{\rm B}T=1,\hspace{1.5cm}\tilde{b}^2=3,
\end{eqnarray}

where $\gamma$ is the effective friction coefficient per charged
monomer. Ultimately, we are interested in how fluctuations in the
charge density decay in space and over time. In terms of the annealed
variables ${\bf R}_a(\tau)$ and the quenched variables
$\theta_a(\tau)$, the charge density is given by $e\hat{\psi}({\bf
r},t)$, where\footnote{The hat on $\hat\psi$ denotes the dependence on
the annealed variables ${\bf R}_a(\tau)$, and the quenched variables
$\theta_a(\tau)$.  This notation should not be confused with the
notation in the MSR-formalism, where the hat indicates conjugated
variables (see appendix A). We stick to these notations because of convention.}

\begin{equation}\label{11}
\hat{\psi}({\bf r},t)=\sum_a \int {\rm d}\tau\hspace{0.05cm}\theta_a(\tau)\hspace{0.05cm} \delta({\bf r}-{\bf R}_a(\tau)).
\end{equation} 

In terms of $\hat{\psi}$, the Fourier transform of the dynamic charge
density correlation function is given by

\begin{equation}\label{12}
g_\theta({\bf k},t)=\frac{e^2}{V} \left< \hat{\psi}(-{\bf k},t)\hat{\psi}({\bf k},t)\right>,
\end{equation}
 
where the brackets denote an average over the annealed variables. Note
that the correlation function depends explicitly on the disorder. The
calculation is simplified considerably by the assumption that the
correlation function is self-averaging, which means that
$g_\theta({\bf k},t)=g({\bf k},t):=\overline{g_{\theta'}({\bf k},t)}$
with probability $1$, where the bar denotes an average over the
quenched variables $\theta'$. This means that in order to find $g_\theta({\bf k},t)$ for a
representative $\theta$ drawn from the Gaussian probability distribution
Eq. (\ref{8}) (which is what we are after), it suffices to calculate the average
of this quantity over the disorder.

We will calculate $g({\bf k},t)$ by means of the Lagrange version of
the MSR (Martin-Siggia-Rose) formalism. In the appendix we have
provided a brief derivation of the MSR functional. It is convenient to
introduce an external field $h({\bf k},t)$ that couples to
$\hat{\psi}({\bf k},t)$. The dynamic charge correlation function can
then be found by differentiating the logarithm of the MSR functional
$Z[h]$ twice with respect to $h$. In principle, the average over the
quenched disorder should be performed {\it after} the differentiation,
but since $Z[h=0]=1$ is independent of the quenched variables (see the
appendix), the order of operations can be interchanged. This is a great
simplification, and it makes the MSR formalism especially suitable for
systems with quenched disorder.

Starting from Eqs. (\ref{A9}) and (\ref{A10}), and substituting the
effective Hamiltonian Eq. (\ref{9}) for the interaction energy ,
one arrives at the expression for the MSR functional. Since analogous
calculations have been published before (see in particular Ref. \cite{Fred})
we will not present the details, but just give the result. During the derivation, the following fields arise:

\begin{eqnarray}\label{13}
\hat\psi_1({\bf k},t)&=&\sum_a \int \hspace{-0.1cm}{\rm
d}\tau\hspace{0.07cm} \theta_a(\tau)\hspace{0.07cm}{\rm exp}\left[
i\hspace{0.02cm}{\bf k}\cdot{\bf R}_a(\tau,t)\right]\nonumber\\ \hat\psi_2({\bf
k},t)&=&\sum_a \int \hspace{-0.1cm}{\rm d}\tau\hspace{0.07cm}
\theta_a(\tau)\hspace{0.07cm}{\bf k}\cdot{\bf
\hat{R}}_a(\tau,t)\hspace{0.07cm}{\rm exp}\left[ i\hspace{0.02cm}{\bf k}\cdot{\bf
R}_a(\tau,t)\right].
\end{eqnarray}

In the
quadratic approximation, the final disorder averaged expression for $Z[h]$ is

\begin{eqnarray}\label{14}
Z[h]\propto \int {\cal D} \psi_1 {\cal D} \psi_2 {\cal D} \phi_1 {\cal D} \phi_2
\hspace{0.05cm}{\rm exp} \left[ -{\cal L}+\int_\omega\int_{\bf k}h(-{\bf k},-\omega)
\hspace{0.05cm}\psi_1({\bf k},\omega)\right],
\end{eqnarray}

where the Lagrangian ${\cal L}$ is given by (there are summations over the indices $i,j$)

\begin{eqnarray}\label{15}
{\cal L}&=&-\ell \int_\omega\int_{\bf k}\frac{4\pi}{k^2} \psi_1(-{\bf
k},-\omega)\hspace{0.05cm}\psi_2({\bf k},\omega)-i\int_\omega\int_{\bf k} \phi_i(-{\bf k},-\omega)\hspace{0.05cm}
\psi_i({\bf k},\omega) \nonumber\\
&&+\frac{c}{2}\int_\omega\int_{\bf k}V_{ij}(-{\bf k},-\omega)
\psi_i(-{\bf k},-\omega)\hspace{0.05cm}\psi_j ({\bf k},\omega).
\end{eqnarray}

$c=n_{\rm{p}}N/V$ is the density of charged monomers. The
integral measures are defined by

\begin{eqnarray}\label{16}
\int_\omega := \frac{1}{2\pi}\int {\rm d}\omega\hspace{2cm}
\int_{\bf k} := \frac{1}{(2\pi)^3}\int {\rm d}^3{\bf k}\hspace{2cm}.
\end{eqnarray}

Although in the full expression for the functional $Z[h]$ there are
more fields present, in the quadratic approximation these fields
couple neither to $\psi_i$, nor to $\phi_i$, and can therefore be
omitted. One consequence of this is that in this approximation the
hydrodynamic interactions do not have an influence on the dynamic
charge density correlation function. Therefore, our results are only
valid on length scales larger than the hydrodynamic screening length.
The integrals over $\psi_i$ and $\phi_i$ in Eq. (\ref{14}) are Gaussian and can be calculated
explicitly, after which $g({\bf k},t)$ can be obtained by differentiating the result
twice with respect to the external field $h$. The result is

\begin{eqnarray}\label{17}
g({\bf k},\omega)=\frac{cV_{11}({\bf k},\omega)}
{\left( 1-\kappa^2k^{-2}V_{12}({\bf k},\omega)\right)
\left( 1-\kappa^2k^{-2}V_{21}({\bf k},\omega)\right)},
\end{eqnarray}

where $\kappa^2:=4\pi \ell c$ is the Debye-H$\ddot{\rm u}$ckel expression \cite{Debye} for the inverse square
screening length. The functions $V_{ij}$ are given by

\begin{eqnarray}\label{18}
V_{11}({\bf k},t)&=&\frac{1}{N}\int {\rm d}\tau\left<{\rm exp}\left[ i{\bf k}\cdot\left(
{\bf R}(\tau,t)-{\bf R}(\tau,0)\right)\right]\right>_0\nonumber\\
V_{12}({\bf k},t)&=&\frac{1}{N}\int {\rm d}\tau\left<{\bf k}\cdot{\bf\hat{R}}(\tau,t){\rm exp}\left[ i{\bf k}\cdot\left(
{\bf R}(\tau,t)-{\bf R}(\tau,0)\right)\right]\right>_0\nonumber\\
V_{21}({\bf k},t)&=&-\frac{1}{N}\int {\rm d}\tau\left<{\bf k}\cdot{\bf\hat{R}}(\tau,0){\rm exp}\left[ i{\bf k}\cdot\left(
{\bf R}(\tau,t)-{\bf R}(\tau,0)\right)\right]\right>_0\nonumber\\
V_{22}({\bf k},t)&=&-\frac{1}{N}\int {\rm d}\tau\left<{\bf k}\cdot{\bf\hat{R}}(\tau,t){\bf k}\cdot{\bf\hat{R}}(\tau,0){\rm exp}\left[ i{\bf k}\cdot\left(
{\bf R}(\tau,t)-{\bf R}(\tau,0)\right)\right]\right>_0.\nonumber\\
\end{eqnarray}

The average $\left<\cdots\right>_0$ is calculated with respect to the single-chain Rouse
Lagrangian ${\cal L}_0$, which is given by

\begin{eqnarray}\label{19}
{\cal L}&=&\int_\omega\int{\rm d}\tau{\bf\hat{R}}(\tau,-\omega)\cdot{\bf\hat{R}}(\tau,\omega)+\int_\omega\int{\rm d}\tau \omega{\bf\hat{R}}(\tau,-\omega)\cdot{\bf R}(\tau,\omega)\nonumber\\
&&-i\int_\omega\int{\rm d}\tau{\bf\hat{R}}(\tau,-\omega)\cdot\frac{\partial^2{\bf R}(\tau,\omega)}{\partial\tau^2}.
\end{eqnarray}

Before starting to calculate $V_{ij}({\bf k},t)$, it is useful to determine what are the
relevant length and time scales. As discussed in the introduction, we
are interested in the processes occurring at length scales of the
order of the blob size $\xi$. As is usual for the semi-dilute regime, the
blob size reaches a finite limit when the chain length approaches
infinity. This limiting value is reached once the chain length exceeds
the value necessary for an isolated chain to collapse into a
globule. It follows that for the processes occurring at length scale $\xi$
the limit $N\rightarrow\infty$ is meaningful, and approached easily in experimental
situations. Therefore, we can safely assume that the wave vectors of
interest satisfy $k\sim\xi^{-1}>>N^{-\frac{1}{2}}$, which will simplify the calculations. Considering
the time scale, it is clear that the processes occurring at length
scale $\xi$ are much faster than those occurring at length scale $N^{\frac{1}{2}}$.
Therefore, in the calculation of $V_{ij}({\bf k},t)$ we can restrict ourselves to times
satisfying $t<<t_R$, where $t_R$ is the Rouse time \cite{Doi} of a single chain. In this
wave vector and time regime it is possible to find explicit and simple
expressions for $V_{ij}({\bf k},t)$. Those readers who are interested in the calculation
leading to Eq. (\ref{20}) are advised to study the appendix of Ref. \cite{Fred},
where similar calculations are worked out in detail. The result is

\begin{eqnarray}\label{20}
V_{11}(k,t)&=&{\rm exp}\left[ -k^2|t|^{1/2}\right]\nonumber\\
V_{12}(k,t)&=&-\theta(-t)\frac{k^2}{|t|^{1/2}}{\rm exp}\left[ -k^2|t|^{1/2}\right]\nonumber\\
V_{21}(k,t)&=&-\theta(t)\frac{k^2}{|t|^{1/2}}{\rm exp}\left[ -k^2|t|^{1/2}\right]\nonumber\\
V_{22}(k,t)&=&0,
\end{eqnarray}

where $\theta(t)$ is the Heaviside step function. We omitted numerical
constants of order unity. Although it is possible to find explicit
expressions for the Fourier transforms $V_{ij}(k,\omega)$, the
resulting expression for $g(k,\omega)$ would be rather
complicated. Instead, it is much more useful to derive a transparent,
albeit approximate, expression for $g(k,\omega)$. An additional
advantage is that this simplified expression can easily be inverse
Fourier transformed with respect to the frequency. Consider the
frequency regime $\omega>>k^4$. Further on we will see that the characteristic 
frequency at wave vector $k$ satisfies this criterion. By substituting $z=-i\omega t$,
distorting the integration contour back to the real axis, and
expanding the integrand in powers of $zk^4/\omega$ \cite{Fred}, one arrives at
the following approximate expressions for $V_{ij}(k,\omega)$:

\begin{eqnarray}\label{21}
|\omega|>>k^4&\Rightarrow &\left\{ 
\begin{array}{l}
V_{11}(k,\omega)\approx|\omega|^{-3/2}k^2\\
V_{12}(k,\omega)\approx\left\{
\begin{array}{l}
(-1+i)|\omega|^{-1/2}k^2\hspace{1cm}\omega>0\\
(-1-i)|\omega|^{-1/2}k^2\hspace{1cm}\omega<0.
\end{array}
\right.
\end{array}
\right.
\end{eqnarray}

Combining Eqs. (\ref{17}) and (\ref{21}) one obtains

\begin{eqnarray}\label{22}
g(k,\omega)\propto\frac{k^2}{\ell |\omega|^{1/2}\left[\left(1+|\omega\kappa^{-4}|^{1/2}\right)^2+1\right]}\hspace{1.5cm}|\omega|>>k^4.
\end{eqnarray}

There are two frequency regimes:

\begin{eqnarray}\label{23}
g(k,\omega)\approx\left\{
\begin{array}{ll}
\ell^{-1}k^2\kappa^{-2}|\omega|^{-1/2}\hspace{1cm}&|\omega|<<\kappa^4\\
\ell^{-1}k^2\kappa^{2}|\omega|^{-3/2}&|\omega|>>\kappa^4.
\end{array}
\right.
\end{eqnarray}

As discussed before, the formalism used can only be expected to
describe the system correctly at length scales larger than the blob
size. Therefore, we will consider Eq. (\ref{23}) only for $k<\kappa$. In that case,
both frequency regimes given in Eq. (\ref{23}) lie within the range of
validity $|\omega|>>k^4$. Doing an inverse Fourier transform with respect to the
frequency we find two regimes:

\begin{eqnarray}\label{24}
c^{-1}g(k,t)\approx k^2\kappa^{-2}\times
\left\{
\begin{array}{lc}
1-|\kappa^4t|^{1/2}\hspace{1cm}&t<\kappa^{-4}\\
|\kappa^4t|^{-1/2}&\kappa^{-4}<t<k^{-4}
\end{array}
\right.
\end{eqnarray}

with the restriction $\xi<k^{-1}<\xi_{\rm e}$ on the wave vector. Note that, in accordance with
the results of Ref. \cite{Joanny1}, the wave vector dependence of the static
correlation function $g(k,t=0)$ coincides with the Debye-H$\ddot{\rm u}$ckel expression $g\propto k^2/\left(k^2+\kappa^2\right)$ in
the regime $k<<\kappa$. Equation (\ref{24}) predicts that charge density fluctuations
decay with a wave vector independent initial rate $\tau\sim\kappa^{-4}\sim \xi^4$. The exponent $4$ is
characteristic for the Rouse model. Since at length scales smaller
than $\xi$ the hydrodynamic interactions are not screened, one would expect
a rate $\tau\sim \xi^3$. The discrepancy is due to the Gaussian approximation
\cite{Fred}. This does not mean that Eq. (\ref{24}) breaks down completely: since at
length scales larger than $\xi$ the hydrodynamic interactions are screened
and the chain {\it is} Rouse-like, Eq. (\ref{24}) is valid for $k<\kappa$, provided that one
inserts the correct basic time scale $\tau$ at the blob level; see the
discussion at the end of the introduction. Defining the dimensionless
wave vector $q:=k/\kappa$ and the dimensionless time $s:=t/\tau$, we obtain for $0.1<q<1$

\begin{eqnarray}\label{25}
c^{-1}g(q\kappa,s\tau)\approx q^2\times
\left\{
\begin{array}{lc}
1-|s|^{1/2}\hspace{1cm}&s<1\\
|s|^{-1/2}&1<s<q^{-4}
\end{array}
\right.
\end{eqnarray}

with

\begin{eqnarray}\label{26}
\tau\sim\frac{\eta\xi^3}{k_{\rm B}T}
\end{eqnarray}									

The expression for $\tau$ reflects the Zimm behavior at small length scales,
whereas the exponent $4$ appearing in the condition $s<q^{-4}$ reflects the Rouse
behavior at longer length scales. The time dependence of the dynamic
charge correlation function is schematically depicted in Fig. 1.

\section{DISCUSSION}

In order to determine the role of the electrostatic interactions in
the expression for the dynamic charge density correlation function,
consider a system in which the electrostatic interactions between the
charges are switched off. Physically, this could be achieved by
immersing the polymer in a concentrated salt solution. Although in
this case the chains would not collapse or phase separate, we will
still assume that the system is semi-dilute, for instance by imposing
a non-zero concentration. In this case, the charge density correlation
function would follow immediately from Eqs. (\ref{17}) and (\ref{20}) by taking $\kappa=0$
. It follows that when the electrostatic interactions are absent, the
charge density correlations decay according to a stretched exponential $g(k,t)\sim k^2\kappa^{-2}{\rm exp}\left[-\left(t/\tau'\right)^{1/2}\right]$ for some $\tau'$
. Considering that the Taylor expansion of this function starts with $1-\left(t/\tau'\right)^{1/2}$,
one sees from Eq. (\ref{25}) that at short times an interacting system
relaxes in qualitatively the same way as a non-interacting system
(though the decay rate is higher), but that at longer times the
relaxation is completely different. We conclude that the appearance of
the power law $g\propto|s|^{-1/2}$ must be due to the Coulomb interactions. In order to
determine the influence of the polymeric bonds, we calculated for
comparison the dynamic charge density correlation function of a salt solution, using
the same formalism and approximations. The result is an exponential
decay of the correlations over time:

\begin{eqnarray}\label{27}
c^{-1}g(k,t)=\frac{k^2}{k^2+\kappa^2}{\rm exp}\left[-\kappa^4t\right]
\end{eqnarray}

The time dependence of Eq. (\ref{27}) is completely different from that of
Eq. (\ref{25}), meaning that the presence of the polymeric bonds has
a large influence on the relaxation of charge density
fluctuations. 

The dynamic correlation function for the {\it total}
density is qualitatively different from that for the charge
density. As an illustration, consider the correlation function
for a semi-dilute homopolymer solution, which has been
calculated within the quadratic approximation in Ref. \cite{Fred}. On
length scales that are small compared to the radius of
gyration of the chains, but large compared to the correlation
length $\xi$, the fluctuations in the total density decay
{\it exponentially} with time (see Eq. (3.19) in Ref. \cite{Fred}).
Experimentally, it would be interesting to test the existence
of the two frequency regimes in Eq. (\ref{23}) by means of
scattering experiments.

\begin{appendix}
\section{}

In this appendix we present a brief derivation of the
Martin-Siggia-Rose functional \cite{MSR}. We will follow the method developed in Refs. 
\cite{Graham,Janssen}. Consider a system of $N$ interacting point
particles in solution. Let $n=1,\cdots,N$ number the particles, let $\alpha=1,2,3$ denote a
Cartesian coordinate, and let $R_{n\alpha}$ be the $\alpha-$component of the position of
particle $n$. Let ${\bf R}$ denote the $3N-$dimensional vector with components $R_{n\alpha}$, and
let $U\left[{\bf R}\right]$ denote the interaction energy. In case that ${\bf R}$ represents the coarse
grained conformation of a polymer chain, entropic contributions have
to be taken into account and one should replace $U$ by an effective
Hamiltonian ${\cal H}$. The time evolution of the probability density $P\left[{\bf R},t\right]$ is
governed by the Fokker-Planck equation \cite{Doi,vanKampen}

\begin{eqnarray}\label{A1}
\frac{\partial P}{\partial t}=\nabla\cdot{\bf L}\cdot\left[k_{\rm B}T\nabla P+P\nabla U\right],
\end{eqnarray}

where $\nabla$ represents the vector with components
$\partial/\partial R_{n\alpha}$, and dots denote inner products. The
mobility matrix ${\bf L}\left[{\bf R}\right]$ is given by \cite{Doi}

\begin{eqnarray}\label{A2}
L_{n\alpha,m\beta}&=&H_{\alpha\beta}\left({\bf R}_n-{\bf R}_m\right)\hspace{1.5 cm}n\not=m\nonumber\\
L_{n\alpha,n\beta}&=&\frac{\delta_{\alpha\beta}}{\gamma}.
\end{eqnarray}

$H_{\alpha\beta}\left({\bf r}\right)$ is the Oseen tensor (see Ref. \cite{Doi}, appendix 3.III) describing the
hydrodynamic interaction, ${\bf R}_n$ denotes the position of particle $n$, and $\gamma$ is
the friction coefficient per particle. In order to derive the
Martin-Siggia-Rose functional it is convenient to switch first from
the Fokker-Planck equation to the equivalent Langevin equation
\cite{vanKampen}. In the Stratonovich interpretation it is given by \cite{Graham}

\begin{eqnarray}\label{A3}
\frac{\partial{\bf R}}{\partial t}=-{\bf L}\cdot\nabla U+k_{\rm B}T\nabla\cdot{\bf L}-\frac{1}{2}{\bf g}\cdot\left(\nabla\cdot{\bf g}\right)+{\bf g}\cdot\mbox{\boldmath$\zeta$}.
\end{eqnarray}

The stochastic force $\zeta$ is a white Gaussian noise with first two moments

\begin{eqnarray}\label{A4}
\left<\mbox{\boldmath$\zeta$}\left(t\right)\right>=0\hspace{1cm}\left<\mbox{\boldmath$\zeta$}\left(t\right)\mbox{\boldmath$\zeta$}\left(t'\right)\right>
={\bf I}\delta\left(t-t'\right).
\end{eqnarray}

${\bf I}$ is the $3N\times 3N$ identity matrix. The matrix ${\bf g}$ is related to the mobility matrix ${bf L}$ by \cite{Graham}

\begin{eqnarray}\label{A5}
{\bf g}\cdot{\bf g}^{\rm T}=2k_{\rm B}T{\bf L}.
\end{eqnarray}

Equation (A5) leaves some freedom in the choice of ${\bf g}$, which can be used
to impose the condition \cite{Graham}

\begin{eqnarray}\label{A6}
\nabla\cdot{\bf g}=0.
\end{eqnarray}

Using this, the Langevin equation simplifies to (it can be shown that $\nabla\cdot{\bf L}=0$)

\begin{eqnarray}\label{A7}
\frac{\partial{\bf R}}{\partial t}=-{\bf L}\cdot\nabla U+{\bf g}\cdot\mbox{\boldmath$\zeta$}+{\bf\hat{h}}.
\end{eqnarray}

The external force $\hat{h}_i(t)$ working on particle $i$ is introduced in order
to be able to calculate response functions. This force will be set
to zero afterwards. In Refs. [5,6] it is
worked out in detail how one can derive, starting from a Langevin
equation, the expression for the path probability distribution
${\cal P}\left[{\bf R}(t)\right]$. By introducing the so-called conjugate
field ${\bf \hat{R}}(t)$, which is rather straightforward in the
Lagrange formalism \cite{Janssen}, one finds the expression

\begin{eqnarray}\label{A8}
{\cal P}\left[{\bf R}(t),{\bf\hat{h}}\right]=\int{\cal D}{\bf \hat{R}}\left(t\right){\rm exp}
\left[-{\cal L}[{\bf R},{\bf \hat{R}}]+\int{\rm d}t{\bf\hat{h}}(t)\cdot i{\bf\hat{R}}(t)\right],
\end{eqnarray}

where the ``Lagrangian'' ${\cal L}[{\bf R},{\bf \hat{R}}]$ is given by

\begin{eqnarray}\label{A9}
{\cal L}[{\bf R},{\bf \hat{R}}]=\int{\rm d}t\left[k_{\rm B}T{\bf \hat{R}}\cdot{\bf L}\cdot
{\bf\hat{R}}+i{\bf\hat{R}}\cdot\left({\bf\dot{R}}+{\bf L}\cdot\nabla U\right)\right].
\end{eqnarray}

The Martin-Siggia-Rose functional is defined by integrating the path
probability over all possible evolutions ${\bf R}(t)$ of the system, in the 
presence of the external fields ${\bf\hat{h}}$ and ${\bf h}$, where the field
${\bf h}$ couples to ${\bf R}$. One obtains

\begin{eqnarray}\label{A10}
Z[{\bf h},{\bf\hat{h}}]=\int{\cal D}{\bf R}\int{\cal D}{\bf \hat{R}}\left(t\right){\rm exp}
\left[-{\cal L}[{\bf R},{\bf \hat{R}}]+\int{\rm d}t{\bf h}(t)\cdot{\bf R}(t)+
\int{\rm d}t{\bf \hat{h}}(t)\cdot i{\bf \hat{R}}(t)\right].
\end{eqnarray}

It follows from Eq. (\ref{A8}) that the correlation and response functions of the system
described by the Langevin equation (\ref{A7}) can be 
obtained from the MSR-functional Eq. (\ref{A10}) by differentiation with respect to
the external fields. However,
the continuum expression for $Z$, as it is given here,
is ill-defined \cite{Langouche}. For instance, it is not possible to determine the
value of the equal-time response function $\left<{\bf R}(t)i{\bf\hat{R}}(t)\right>$. 
Retracking the derivation
of Eq. (\ref{A10}) one finds that the discretization underlying the integrals
over time is such that the equal-time response functions are
identically zero \cite{Langouche}. This extra information is sufficient to remove
all ambuigity from Eq. (A10).

We discuss briefly the adequacy of the Langevin equation Eq. (\ref{A7}) to
describe the dynamics of solutions. It has been argued \cite{Oeno} that
Eq. (\ref{A7}) is inconsistent, in the sense that whereas the thermal
fluctuations of the particles are taken into account, the thermal
fluctuations of the solvent velocity field are not (the Oseen tensor
gives the {\it average} solvent velocity as a function of the forces). In
order to obtain the same level of description for both the particles
and the solvent, Oono and Freed \cite{Oeno} introduced a set of coupled
Langevin equations. However, it seems to us that the only difference
between these Oono-Freed equations and Eq. (\ref{A7}) lies in the neglect of
the solvent inertia in the latter, for the following reason. It was shown in Ref. \cite{Fred} that if
one derives the MSR functional from the Oono-Freed
equations, the velocity and its conjugate appear quadratic in the
Lagrangian, and so they can be integrated out explicitly. The
resulting functional is identical to the one obtained from Eq. (\ref{A7}),
provided that one takes the solvent density to be zero. Since the
effects of solvent inertia on the dynamics of polymer solutions are
negligible anyway, we conclude that Eq. (\ref{A7}) is equivalent to the
Oono-Freed equations. As an illustration of the irrelevancy of the
solvent inertia, consider the Zimm model. The characteristic frequency $\omega$
of fluctuations with wave vector $k$ is given by \cite{Doi} $\omega=k_{\rm B}Tk^3/6\pi\eta$, where 
$\eta$ is the
solvent viscosity. It follows from the Navier-Stokes equation that the
solvent inertia is negligible if $\rho\omega<<\eta k^2$, where $\rho$ is the solvent
density. Taking the viscosity and the density of water one finds that
the effects of the solvent inertia are negligible on length scales $L>>10^{-13}m$
. Nevertheless, if one is interested in the correlation and response
functions of the solvent velocity field, the Oono-Freed equations are
certainly usefull \cite{Fred}.
\end{appendix}
\\
ACKNOWLEDGMENTS

Hindrik Jan Angerman is grateful to the NWO (Netherlands Organization
for Scientific Research) for financial support in the form of a
NATO-Science Fellowship.

FIGURE CAPTION

The dynamic charge density correlation function $g(q,s)$ as a function of 
the dimensionless time $s:=t/\tau$ for fixed dimensionless wave vector $0.1<q:=k\xi<1$,
where $\xi$ is the blob size, and $\tau$ is the characteristic time scale at the blob
level. 

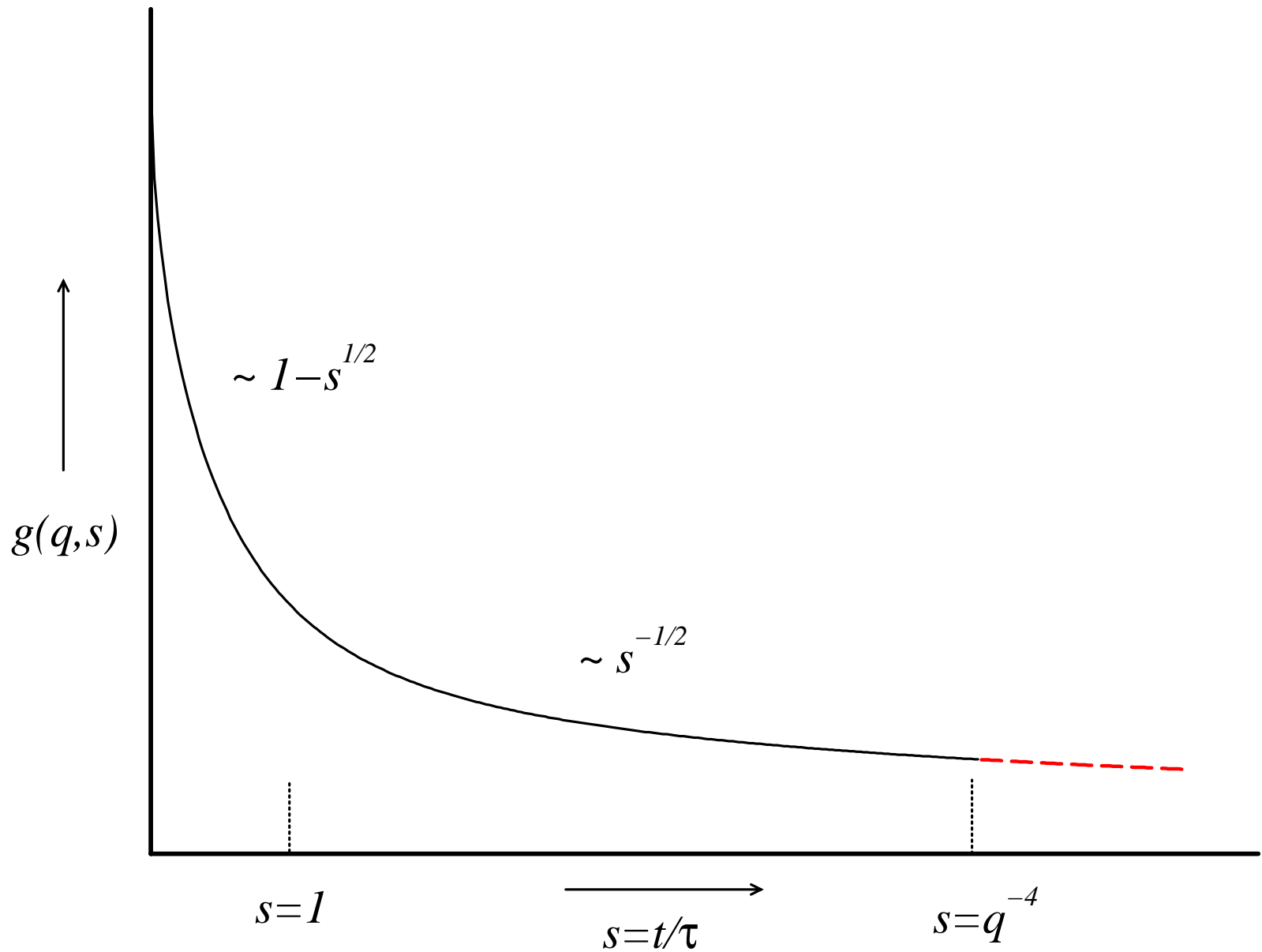
\end{document}